\documentclass[a4paper]{article}
\usepackage{epsfig,amssymb,amsmath}

\def\beq{\begin{eqnarray}}
\def\eeq{\end{eqnarray}}
\def\bsp{\begin{split}}
\def\esp{\end{split}}

\def\inl{\left\langle}
\def\inr{\right\rangle}
\def\dt{\partial_t}

\def\d{\mathrm{d}}

\newcommand{\mbold}[1]{\mbox{\boldmath{\ensuremath{#1}}}}

\begin{document}

\title{\textbf{Brane Waves}}
\author{\textbf{Alan A. Coley}\thanks{%
aac@mathstat.dal.ca}~ and\ \textbf{Sigbj\o rn Hervik}\thanks{%
herviks@mathstat.dal.ca} \\
\\
Department of Mathematics and Statistics,\\
Dalhousie University, \\
Halifax, Nova Scotia, Canada B3H 3J5}
\date{\today}
\maketitle

\begin{abstract}

In brane-world cosmology gravitational waves can 
propagate in the higher dimensions (i.e., in the `bulk').
In some appropriate regimes the
bulk gravitational waves may be approximated by plane waves.
We systematically study five-dimensional gravitational waves that are
algebra\-ically special and of type N.  
In the most physically relevant case the projected non-local stress tensor
on the brane is formally equivalent to the energy-momentum tensor of a null fluid. 
Some exact solutions are studied to 
illustrate the features of these branes; in particular, we show explicity that any plane wave brane can be embedded into a 5-dimensional Siklos spacetime. 
More importantly, it is possible that in some appropriate regime the bulk can be approximated by gravitational
plane waves and thus may act
as initial conditions for the gravitational field in the bulk (thereby enabling
the field equations to be integrated  on the brane).

\end{abstract}

\section{Introduction}

In brane-world cosmology, matter fields and gauge
interactions are confined to a four-dimensional brane
moving in a higher-dimensional ``bulk" spacetime. This paradigm is motivated 
by string and M-theory; in particular, generalized
Randall-Sundrum-type models \cite{randall} are relatively simple
phenomenological five-dimensional (5D) models which capture some of the essential features of the
dimensional reduction of Ho$\check{\mbox{r}}$ava-Witten theory~\cite{Horava}.
In a recent analysis of the asymptotic dynamical evolution of perfect fluid
brane-world cosmological models close to the initial singularity,
it was found that for an appropriate range of the
equation of state parameter an isotropic singularity
is a past-attractor \cite{COLEY}.
It was subsequently argued that the initial cosmological singularity
is isotropic in brane-world cosmological models.

The 5D field equations are 
\begin{equation}\label{5}
^{(5)}G_{ab}=-\Lambda_5\, ^{(5)}g_{ab} +
\kappa_5^2\,^{(5)}T_{ab}\,, ~~\Lambda_5=-\frac{6}{\ell^2}\,.
\end{equation}
Here, $a,b=0,...,4$, $\Lambda_5$ and $g_{ab}$ is the 5D cosmological constant and metric, respectively. 
The projected field equations on the
brane are ~\cite{SMS}:
\begin{equation}\label{modfe}
G_{\mu\nu}=-\Lambda g_{\mu\nu}+ \kappa^2 T_{\mu\nu} +
6{\frac{\kappa^2}\lambda}{\cal S}_{\mu\nu} - {\cal E}_{\mu\nu}\, ,
\end{equation}
where $\mu,\nu$ are brane indices (i.e. $\mu,\nu=0,..,3$) $\Lambda = {\Lambda_5 /2} + \lambda^2 \kappa_5^4/12$ and
$\kappa^2= {\lambda }\kappa_5^4/6$. The term ${\cal S}_{\mu\nu}$
is quadratic in $T_{\mu\nu}$ and dominates at high energies
($T_{00}=\rho>\lambda$), and the
five-dimensional Weyl tensor is felt on the brane via its
projection, $\mathcal{E}_{ab}=C_{acbd}n^cn^d$, where $n^a$ is the unit normal vector to the brane. 
There may also be terms that arise from from 5D sources in
the bulk other than the vacuum energy $\Lambda_5$, such as a bulk
dilaton field. In general, in the 
4-dimensional picture the conservation equations do not
determine all of the independent components of ${\cal E}_{\mu\nu}$
on the brane (and a complete higher-dimensional analysis,
including the dynamics in the bulk, is necessary) \cite{maartens}.

In these models the gravitational field can also
propagate in the extra dimensions (i.e., in the `bulk').
For example, there might occur thermal radiation of bulk gravitons \cite{Lang}.
In particular, at sufficiently
high energies particle interactions
can produce 5D gravitons which are emitted into the bulk. Conversely,
in models with a bulk black hole, there may be gravitational waves
hitting the brane. At sufficiently large distances from the black hole
these gravitational waves may be approximated as of type N. Alternatively, if the brane has low energy initially, energy can be transferred onto the brane by bulk particles such as gravitons; an equilibrium is expected to set in once the brane energy density reaches  a limiting value. (For an alternative approach  see, e.g., \cite{KKTTZ}.) In this paper
we shall study the consequences of assuming  that in some appropriate regimes the
bulk gravitational waves can be approximated by plane waves.


\section{An example of a Type N bulk with a plane wave brane}
\label{sect:AdSWaves}

First, let us  consider an example of an exact 5D solution of type N with a 
negative cosmological constant. This particular example will give us some hints what 
we can expect from the brane-world analysis. In particular, the example makes it possible 
to interpret some of the exact solutions on the brane discussed later. 

Consider the Siklos metric \cite{Siklos} (where we have dropped an overall scaling)
\beq
\d s^2=\frac{1}{z^2}\left(2\d v\d u+H(u,x,y,z)\d u^2+\d x^2+\d y^2+\d z^2\right), \quad z>0. 
\label{eq:Siklos}\eeq
This metric is an Einstein space (i.e., $R_{ab}=\Lambda g_{ab}$) 
if the function $H$ solves the equation:
\beq
\left(\frac{\partial^2}{\partial x^2}+\frac{\partial^2}{\partial y^2}+\frac{\partial^2}{\partial z^2}-\frac 3z\frac{\partial}{\partial z}\right)H(u,x,y,z)=0.
\label{eq:BoxH}\eeq
These solutions describe gravitational waves propagating in a negatively curved Einstein space. Hence, these metrics generalise the AdS$_5$ spaces to solutions of the Einstein equation with a negative cosmological constant containing gravitational waves. These are exactly the type of models we want to investigate for the bulk. 

To see  how these can generate anisotropic stresses in the bulk, 
we can calculate the Weyl tensor of the above solutions. In general 
the Weyl tensor has the following non-vanishing components (in a coordinate basis):
\beq
{C_{uiuj}}&=& -\frac 1{2z^2}\frac{\partial^2}{\partial x^i\partial x^j}H, \nonumber \\
C_{uiui}&=&-\frac{1}{6z^2}\left(2\frac{\partial^2}{\partial {x^i}^2}-\frac{\partial^2}{\partial {x^j}^2}-\frac{\partial^2}{\partial {x^k}^2}\right)H,
\eeq
where $(i,j,k)$ is a permutation of $(x,y,z)$. Since $\mathcal{E}_{ab}=C_{acbd}n^cn^d$, 
the Weyl tensor can induce anisotropic stresses on the 
brane. For example, the Weyl tensor corresponding to the gravitational mode
$H= x^2-y^2$, yields the bulk stresses on the brane given by
$C_{uxux}=-C_{uyuy}=-\frac{1}{z^2}$.
We also have oscillatory modes  by choosing, for example,
\beq
H(u,x,y,z)=(A+Cz^4)(x^2-y^2)\cos(u+\varphi_1)+(D+Bz^4)\cos(u+\varphi_2),
\eeq
where $A,~B,~C,~D,~ \varphi_1$ and $\varphi_2$ are arbitrary constants. 
We will later see that similar modes naturally enter the analysis of the brane. 

Let us, for the sake of illustration, consider a simple exact 5-dimensional solution of a brane of this type. We consider a case in which a brane is embedded in the 5-dimensional solution eq.(\ref{eq:Siklos}). The brane is located at  $z=z_0$ with normal unit vector ${\bf n}=z\partial/\partial z$. Note that if 
\beq
\left.\frac{\partial H}{\partial z}\right|_{z=z_0}=0,
\label{eq:Junction}\eeq 
we obtain for the extrinsic curvature $K_{\mu\nu}\propto g_{\mu\nu}$, where $g_{\mu\nu}$ is the induced metric on the brane. Hence, we see from the junction conditions that there are some vacuum solutions\footnote{Also note that a propagating electromagnetic wave on the brane can support a non-zero $\left.\frac{\partial H}{\partial z}\right|_{z=z_0}$.} by suitably tuning the brane tension $\lambda$ in the usual way. The induced metric on the brane can be written 
\beq
\d s^2_{\text{brane}}=2\d v\d u+\mathcal{H}(u,x,y)\d u^2+\d x^2+\d y^2,
\eeq
and hence, is a plane wave. Using eq.(\ref{eq:BoxH}) and the junction conditions, the function $\mathcal{H}(u,x,y)\equiv H(u,x,y,z_0)$ obeys
\beq
\left(\frac{\partial^2}{\partial x^2}+\frac{\partial^2}{\partial y^2}\right)\mathcal{H}(u,x,y)=-\left.\frac{\partial^2}{\partial z^2}H(u,x,y,z)\right|_{z=z_0}
\label{eq:BoxcalH}.
\eeq
These brane solutions are therefore not the usual vacuum solutions; they are in general sourced by the projected  Weyl tensor of the bulk (see e.g. \cite{Kramer}). The gravitational waves in the bulk propagate along the brane and induce, via the projected Weyl tensor, a non-local anisotropic stress. This stress mimics the effect of an electromagnetic wave travelling on the brane at the speed of light. 

We note that \emph{any} plane wave brane can be embedded (locally) in a Siklos spacetime. Given $\mathcal{H}(u,x,y)$, a solution can be found by choosing
\beq
H(u,x,y,z)=\mathsf{D}(z,\nabla_2^2)\mathcal{H}(u,x,y), \quad \nabla_2^2\equiv \frac{\partial^2}{\partial x^2}+\frac{\partial^2}{\partial y^2},
\eeq
where $\mathsf{D}(z,\nabla_2^2)$ is the operator 
\beq
\mathsf{D}(z,\nabla_2^2)&=&\sum_{i=0}^{\infty}F_i(z)\left(\nabla_2^2\right)^i,
\eeq
and $F_i(z)$ are defined iteratively: 
\beq
F_0(z)=1, \quad F_{i+1}(z)=-\int_{z_0}^zz^3\left(\int_{z_0}^z \frac{1}{z^3}F_i(z)\d z\right)\d z.
\eeq
The brane is located at $z_0$ and $H$ is chosen such that eq.(\ref{eq:Junction}) is fulfilled. In the special case where $\mathcal{H}(u,x,y)$ is a polynomial in $x$ and $y$, the above sum will terminate and  $H$ will only contain a finite number of terms. Furthermore, by choosing the function $\mathcal{H}(u,x,y)$ appropriately, we note that the above example can include branes of Bianchi type III, IV, V, VI$_h$ and VII$_h$. The relationship between brane and bulk can be further studied in a full 5-dimensional setting. 

\section{A general type N bulk} 

Let us now be more systematic and assume that the 5D bulk
is algebraically special and of type N.  This puts a constraint on the 5D Weyl tensor which
makes it possible to deduce the form of the non-local stresses from a brane point of view.
For a 5D type N spacetime there exists a frame $\ell_a$, $\tilde{n}_a$, $m^i_a$, $i=2,3,4$  
such that \cite{classification}:
\beq
C_{abcd}=4C_{1i1j}\ell_{\{a}m^i_b\ell_cm^j_{d\}}.
\eeq
The frame $\ell_a$, $\tilde{n}_a$, $m^i_a$ (the frame vector $\tilde{n}^a$ is not to be confused with the brane normal vector $n^a$) is defined via the only non-zero contractions: 
\[ \ell_a\tilde{n}^a=\ell^a\tilde{n}_a=1, \quad m^i_am^{ja}=\delta^{ij}. \] 
The ``electric part'' of the Weyl tensor, $\mathcal{E}_{ab}=C_{acbd}n^cn^d$, where $n^c$ is the normal vector on the brane, can for the type N bulk be written as
\beq
\mathcal{E}_{ab}=C_{1i1j}\left[\ell_a(m_c^in^c)-m^i_a(\ell_cn^c)\right]\left[\ell_b(m_c^jn^c)-m^j_b(\ell_cn^c)\right]. 
\eeq
Furthermore, one can easily check that $\mathcal{E}_{ab}n^b=\mathcal{E}^a_{~a}=0$. Note  that for a type N bulk we also have  
\beq
\mathcal{E}_{ab}\ell^b=0.
\eeq
This can be rewritten using the projection operator on the brane, $\tilde{g}^a_{~b}={g}^a_{~b}-n^an_b$,
\beq
\mathcal{E}_{ac}\tilde{g}^c_{~b}\ell^b=\mathcal{E}_{ab}\hat{\ell}^b=0, \quad \hat{\ell}^b\equiv \tilde{g}^b_{~c}\ell^c.
\eeq
Hence, the vector $\hat{\ell}_b$ is the projection of the null vector $\ell_b$ onto the brane. By contracting this vector with itself, we get
\beq
\hat{\ell}^b\hat{\ell}_b=-(\ell^an_a)^2. 
\eeq
The following analysis splits into two cases, according to whether $\ell^an_a$ equals zero or not:
\begin{enumerate}
\item{} $\ell^an_a=0$: $\hat{\ell}_a=\ell_a$ and null.
\item{} $\ell^an_a\neq 0$: $\hat{\ell}_a$ time-like. 
\end{enumerate}

These cases have to be treated separately and have different interpretations on the brane.
From a 5D point of view they are of the same type, but since the 5D spacetime is
anisotropic the orientation of the brane with respect to $\ell_a$ is of significance.  More
precisely, considering plane-wave spacetimes, the case $\ell^an_a=0$ corresponds to when
the wave propagates parallel to the brane, and in the case $\ell^an_a\neq 0$ the wave hits the
brane.

\subsection{The case $\ell^an_a=0$}

Let us first investigate the consequence of $\hat{\ell}_{\mu}$ being a null vector. 
We note that $\mathcal{E}_{\mu\nu}$, the four-dimensional projected Weyl tensor on the brane,
can be written in this case
\beq
\mathcal{E}_{\mu\nu}=-\left(\frac{6}{\lambda\kappa^2}\right)\epsilon\hat{\ell}_{\mu}\hat{\ell}_{\nu},
\eeq
where $\epsilon$ is some appropriate function. 
Hence, this is formally equivalent to the energy-momentum tensor of a \emph{null} fluid. 
Equivalently we can consider it as the energy-momentum tensor of an extreme 
tilted perfect fluid. Using a covariant decomposition of $\mathcal{E}_{\mu\nu}$ with respect to a preferred time-like vector $u_{\mu}$ being orthogonal to some 3-surface with metric $h_{\mu\nu}$, we have
\beq 
\mathcal{E}_{\mu\nu}=-\left(\frac{6}{\lambda\kappa^2}\right)\left[\mathcal{U}\left(u_{\mu}u_{\nu}+\frac 13 h_{\mu\nu}\right)+\mathcal{P}_{\mu\nu}+2\mathcal{Q}_{(\mu}u_{\nu)}\right].
\eeq
The non-local energy terms are thus given by \cite{maartens}:
\beq
\mathcal{U}&=&\epsilon(\hat{\ell}_{\nu}u^{\nu})^2, \nonumber \\
\mathcal{Q}_{\mu}&=&\epsilon(\hat{\ell}_{\nu}u^{\nu})\hat{\ell}_{\mu}, \nonumber \\
\mathcal{P}_{\mu\nu}&=&\epsilon\hat{\ell}_{\langle\mu}\hat{\ell}_{\nu\rangle}.
\eeq
Here, angled brackets $\langle\cdots\rangle$ denotes the projected, symmetric and tracefree part with respect to the metric $h_{\mu\nu}$ of the spatial 3-surfaces. 
The equations on the brane now close and the dynamical behaviour can be analysed
(see later, and \cite{CH}). Note that
\beq
\mathcal{U}\mathcal{P}_{\mu\nu}=\mathcal{Q}_{\mu}\mathcal{Q}_{\nu} - \frac{1}{3}{g}_{\mu\nu}
\mathcal{Q}_{\lambda}\mathcal{Q}^{\lambda},
\eeq
so that in this case $\mathcal{E}_{\mu\nu}$ is determined completely by
$\mathcal{U}$ and $\mathcal{Q}_{\mu}$.

\subsection{The case $\ell^an_a\neq 0$}
In this case the vector $\ell^a$ has a component  orthogonal to the brane. 
This implies that the vector $\hat{\ell}^a$ is time-like. 
This vector lives on the brane, and hence we can set $u^{\mu}$ parallel to 
$\hat{\ell}^{\mu}$. In this frame, the requirement $\hat{\ell}^{\mu}\mathcal{E}_{\mu\nu}=0$ 
implies that we have $\mathcal{U}=0=\mathcal{Q}_{\mu}$ and hence, we can write 
\beq
\mathcal{E}_{\mu\nu}=-\left(\frac{6}{\lambda\kappa^2}\right)\mathcal{P}_{\mu\nu}.
\eeq
This is formally equivalent to a fluid which possesses anisotropic stresses
with no energy density or energy-flux. From a brane point of view it appears as if these 
stresses are superluminal; however, as can be seen from from a 5D point of view, this is 
just an artifact of living on a brane in a higher-dimensional spacetime. The stresses 
\emph{do} have a gravitational origin, namely from gravitational waves in the 5D bulk.

\subsection{Exact solutions describing radiating branes}

Next, we will find some exact solutions in the case $\ell^an_a\neq 0$, which are
very illustrative and describe features of branes that seems to have gone unnoticed in
the literature.
We assume that the brane contains an isotropic fluid, i.e. the energy-momentum tensor takes the form
\beq
T_{\mu\nu}=(p+\rho)u_{\mu}u_{\nu}+p\tilde{g}_{\mu\nu},
\eeq
and that $p=(\gamma-1)\rho$, $\mathcal{Q}_{\mu}=\mathcal{U}=0$. We also assume that the brane has isotropic 
curvature, $\mathcal{R}_{\inl\mu\nu\inr}=0$ (e.g., is of Bianchi type I  
or V). From the propagation equation for 
the non-local dark energy $\mathcal{U}$, the simple constraint
\beq
\sigma_{\mu\nu}\mathcal{P}^{\mu\nu}=0,
\eeq
where $\sigma_{\mu\nu}\equiv \nabla_{\langle\mu}u_{\nu\rangle}$ is the shear, 
is obtained (see \cite{MSS}).
Since both $\sigma_{\mu\nu}$ and $\mathcal{P}_{\mu\nu}$ are symmetric, 
space-like, trace-free and of maximal rank 3, and defining the 
natural inner product, the constraint $\sigma_{\mu\nu}\mathcal{P}^{\mu\nu}=0$ 
implies that $\sigma_{\mu\nu}$ and $\mathcal{P}_{\mu\nu}$ are orthogonal. 
We have the evolution equation
\beq
\dt\sigma^{\mu}_{~\nu}+\Omega^{\mu}_{~\alpha}\sigma^{\alpha}_{~\nu}-\sigma^{\mu}_{~\alpha}\Omega^{\alpha}_{~\nu}+\theta\sigma^{\mu}_{~\nu}=\frac{6}{\kappa^2\lambda}\mathcal{P}^{\mu}_{~\nu},
\eeq
where $\theta=\nabla^{\mu}u_{\mu}$ and 
$\Omega^{\alpha}_{~\nu}$ is antisymmetric and is the rotation tensor with respect to a set of Fermi-propagated axes (recall that an overdot is defined by $\dot{}\equiv u^{\mu}\nabla_{\mu}$).
Writing $\theta=3\dot{a}/a$, the shear equations then give
\beq
\dt(\sigma_{\mu\nu}\sigma^{\mu\nu})+6\frac{\dot{a}}{a}\sigma_{\mu\nu}\sigma^{\mu\nu}=0.
\eeq
This can easily be integrated to give $\sigma_{\mu\nu}\sigma^{\mu\nu}=A^2 a^{-6}$. 
Thus the shear scalar decays withe the usual $a^{-6}$ behaviour and is unaffected by 
the non-zero $\mathcal{P}_{\mu\nu}$. This is simply due to the fact that 
$\mathcal{P}_{\mu\nu}$ is orthogonal to $\sigma_{\mu\nu}$ and thus it cannot 
affect the shear scalar. The generalised Friedmann equation now becomes  \cite{MSS}
\beq
\frac{\dot{a}^2}{a^2}=\frac{\Lambda}{3}+\frac{A^2}{3}\frac{1}{a^6}-\frac{1}{6}\frac{k}{a^2}+\frac{\kappa^2}{3}\frac{\rho_0}{a^{3\gamma}}\left(1+\frac{1}{2\lambda}\frac{\rho_0}{a^{3\gamma}}\right),
\eeq
where, for example,  $k=0$ for Bianchi type I and $k=-1$ for Bianchi type V.
This equation can now easily be solved in quadrature. 
Due to the quadratic terms in the energy-densities arising from brane effects, 
the shear mode is subdominant in the past as long as the isotropic fluid is stiffer than dust.

It is convenient to define two new variables 
\[ X_{\mu\nu}=a^3\sigma_{\mu\nu}, \quad Y_{\mu\nu}=a^3\mathcal{P}_{\mu\nu}.\]
In tensor form, the shear evolution equations are 
\beq
\dt{\sf X}+[\mbold{\Omega},{\sf X}]=\omega{\sf Y},
\eeq
where $\omega=6/(\kappa^2\lambda)$. 
This equation can be integrated in quadrature once we know ${\sf Y}$ (recall that
from a brane point of view, 
there are no evolution equations for ${\sf Y}$). 
We solve the above equations by choosing the particular gauge in which  
$\mbold{\Omega} = 0$ (${\sf X}$ is not diagonal in general), 
so that the axes are Fermi-propagated. A solution can now be written:
\beq
{\sf X}&=&{\sf X}_0\cos\omega t+{\sf Y}_0\sin\omega t, \\
{\sf Y}&=&{\sf Y}_0\cos\omega t-{\sf X}_0\sin\omega t.
\eeq
In this solution it is easy to show that if  ${\sf X}_0$, ${\sf Y}_0$ are 
orthogonal initially, then ${\sf X}$ and ${\sf Y}$ will be orthogonal at all times. 
From the assumption of a periodic ${\sf Y}$, it follows that the function $H$ satisfies
a wave equation as desired.



The above solutions describe brane-worlds where the shear oscillates with a frequency \[
f=\frac{2\pi}{\omega}=\frac{\pi\kappa^2\lambda}3.\] This oscillation is driven by a similar
oscillation in the non-local anisotropic stresses $\mathcal{P}_{\mu\nu}$.  The frequency of
the oscillation is given by the brane-tension and the gravitational constant on the brane
and hence the physics on the brane.  If these
gravitational waves came from a source in the bulk, then that source must somehow know of
the physics on the far-away brane.  Unless there is a very special configuration in the
bulk, then this situation does not seem very likely.  A more likely scenario is that it is
the \emph{brane itself} which is the source of the gravitational radiation.  The brane
oscillates with a given frequency which is dictated by the brane tension and the
gravitational constant on the brane.  This oscillation causes the brane to emit
gravitational radiation into the bulk.  The frequency of the gravitational radiation will
thus carry a distinct signature given by its frequency.  
If the bulk consists solely of an incoming or an outgoing
wave then the type N approximation is valid.
Suppose there is a source in the bulk emitting
radiation with a different characteristic frequency than the brane.  
Then close to the membrane there will be a mixture of incoming and
outgoing waves, unless some sort of equilibrium is established.
This implies that the simple assumption that there exists a preferred wave
vector just outside the brane (and the bulk just outside the brane 
is of type N) breaks down.

\section{The effect of gravitational waves on the dynamical behaviour of the brane}

Finally, we will investigate what kind of effect this type N bulk may have on the cosmological 
evolution of the brane. It is believed that an isotropic singularity is the most likely 
initial starting point for a classical brane-world \cite{COLEY}. For this to be correct, the initial 
singularity has to be stable into the past. Let us assume, therefore, that we are in the regime of 
an isotropic past and the cosmological evolution is dominated by an isotropic perfect 
fluid with equation of state $p=(\gamma-1)\rho$. In this approximation, and assuming the 
case $\ell^an_a=0$, the equations for $\mathcal{U}$ and $\mathcal{Q}_{\mu}$ are
\beq
\dot{\mathcal{U}}+\frac 43\theta\mathcal{U}=0, \quad \quad 
\dot{\mathcal{Q}}_{\mu}+\frac 43\theta\mathcal{Q}_{\mu}=0.
\eeq
For the isotropic singularity, the expansion factor is given by $\theta=1/(\gamma t)$. Hence, defining the expansion-normalised non-local density, $U\equiv\mathcal{U}/\theta^2$, and non-local energy flux, $Q_{\mu}\equiv\mathcal{Q}_{\mu}/\theta^2$ we get
\beq
\dot{U} = \frac{2}{3\gamma}(3\gamma-2)U, \quad \quad
\dot{Q}_{\mu} = \frac{2}{3\gamma}(3\gamma-2)Q_{\mu}.
\eeq 
Hence, the isotropic singularity 
is stable to the past with regards to these stresses if $\gamma>2/3$. 

Similarly, for the case $\ell^an_a=0$ we have a FRW universe to the future with 
$\theta=2/(\gamma t)$, 
and thus 
\beq
\dot{U} = \frac{2}{3\gamma}(3\gamma-4)U, \quad \quad
\dot{Q}_{\mu} = \frac{2}{3\gamma}(3\gamma-4)Q_{\mu}.
\eeq 
This implies that if the isotropic fluid on the brane is stiffer than radiation 
($\gamma=4/3$), then the FRW universe is unstable to the future with respect to the non-local 
stresses. However, since dust has $\gamma=1$, our physical universe is  believed to be stable 
with respect to these non-local stresses. 
\section{Discussion}
These phenomenological results are further supported by a more detailed analysis of the asymptotic
behaviour of two tilting $\gamma$-law fluids in a class of Bianchi type VI$_0$ models 
\cite{CH}.
In particular, the physically relevant case of interest here, namely that the second fluid is a null
fluid or fluid with extreme tilt, is investigated. All equilibrium points are found and their stability
determined, so that the local attractors can be established. It is found that the 
dynamical effects of the projected
Weyl tensor are not significant asymptotically at early and late times.

In brane cosmology the only degrees of 
freedom in the bulk are the higher-dimensional gravitational waves, which propagate 
in the bulk \cite{SMS,Lang}. The problem of initial conditions for these bulk gravitational waves are 
of great importance \cite{BBC}. Indeed, we
need initial conditions for the gravitational field in the bulk to be able
to integrate the field equations and determine the dynamics on the brane.
We have argued that in some situations the bulk
gravitational waves can be approximated as type N plane-wave solutions.
Consequently, perhaps 
an appropriate set of initial conditions is to assume plane behaviour
in a suitable regime of the bulk. This can be achieved by demanding that 
the 5-dimensional bulk is algebraically special and of type N as above.
The effects this type N bulk may have to the cosmological 
evolution of the brane is studied in \cite{CH}.

\subsection*{Acknowledgments} 
We would like to thank Roy Maartens for comments. AC was funded by NSERC and SH was funded by an AARMS and a Killam PostDoctoral Fellowship.

\end{document}